%% file: main_preprint.tex
\documentclass[journal]{IEEEtran}
\usepackage[utf8]{inputenc}
\usepackage[english]{babel}
\usepackage{multicol}               % Kolonner
\usepackage[a4paper, margin = 1in]{geometry}      % Struktur

\usepackage{mathtools} % Matte
\usepackage{amsmath}
\usepackage{graphicx} % Grafikk
\usepackage{float}
\usepackage{fancyhdr} % Header 

\usepackage{comment}
\usepackage[table]{xcolor}
\usepackage{multirow}

\usepackage{caption}
\usepackage{subcaption}

\usepackage[ruled,vlined,linesnumbered]{algorithm2e}

\usepackage{hyperref}
\usepackage{csquotes}
\usepackage{cleveref}
\usepackage{makecell}
\usepackage{multicol}

\renewcommand\IEEEkeywordsname{Keywords}

\title{A Tool for Reliability Assessment of Smart and Active Distribution Systems - RELSAD} %An Open-Source
\author{
\IEEEauthorblockN{Stine Fleischer Myhre\IEEEauthorrefmark{1}, Olav Bjarte Fosso\IEEEauthorrefmark{1}, Poul Einar Heegaard\IEEEauthorrefmark{2}, Oddbjørn Gjerde\IEEEauthorrefmark{3}}
\IEEEauthorblockA{\IEEEauthorrefmark{1}Department of Electric Power Engineering, NTNU\\
\IEEEauthorrefmark{2}Dept. of Information Security and Communication Technology, NTNU  \\
\IEEEauthorrefmark{3}SINTEF Energy Research\\
Trondheim, Norway\\
\{stine.f.myhre, olav.fosso, poul.heegaard\}@ntnu.no; oddbjorn.gjerde@sintef.no}
}
\date{February 2020}

\begin{document}
\maketitle

\begin{abstract}
    With increased penetration of new technology in the distribution systems such as renewable energy resources, flexible resources, and information and communication technology, the distribution systems become more complex and dynamic. The traditional reliability analysis methods do not consider the new components and technology and new considerations need to be taken to address these new changes in the distribution system. This paper presents an open-source reliability assessment tool for smart and active distribution systems (RELSAD). The tool aims to function as a foundation for reliability calculation in the smart and active distribution systems, where these new components and technologies are included. The tool is made as a Python package built up based on an object-oriented programming approach. The method will be illustrated on the IEEE 33-bus system with the inclusion of generation, battery, and ICT. 
    
    %With higher penetration of renewable energy resources, flexible resources such as batteries, and information and communication technology (ICT), the system becomes more active and smarter. T
\end{abstract}

\begin{IEEEkeywords}
    Distribution systems, ICT, Object-oriented programming, Python, Reliability assessment, Smart-grid
\end{IEEEkeywords}

\input{Sections/01Introduction.tex}
\input{Sections/02SystemStructure.tex}

\input{Sections/03SystemFunction.tex}

\input{Sections/04UserInterface.tex}
\input{Sections/06CaseStudy.tex}

\input{Sections/07Discussion.tex}
\input{Sections/08Conclusion.tex}
\input{Sections/09Acknowledgement.tex}

\bibliographystyle{IEEEtran}
\bibliography{main_preprint}

% References: 

\begin{comment}

\end{comment}

\end{document}

%% file: Sections/01Introduction.tex
\section{Introduction}

The electrical power system is under constant change, and some of these changes are observed at the distribution system level. With higher integration of renewable energy resources, flexible end-users, energy storage, and ICT components, the system becomes more complex with bidirectional power and communication flow \cite{prettico2019distribution, ipakchi2009grid}. However, even though the distribution system is moving towards an active system with more similar behavior as the transmission system, the distribution systems are still operated as radial or weakly meshed systems.

The traditional reliability analysis methods for distribution systems do not consider the new components and technologies in the distribution system \cite{escalera2018survey}. Typically, only passive operation of the distribution systems, often systems without any generation, is considered. If generation units are present, they will be disconnected during the whole outage period of the faulted line and not participate in an active way to restore supply to isolated system parts.

%In cases with implemented units in the distribution system, these will be disconnected during    

%They consider the passive operation of the distribution systems, often systems without any production units present. 

Reliability evaluation of radially operated distribution systems is conducted mainly through analytical approaches or simulation. The common goal is to evaluate the impact on the distribution system and the system customers of the possible outage scenarios in the system. The basic calculated reliability parameters are the average failure rate, $\lambda_s$, average outage time, $r_s$, and average annual outage time, $U_s$ for the system. In addition, it is common to calculate interruption indices to give a better representation of the system behavior. 

In \cite{billinton1992reliability}, multiple analytical approaches for evaluating the reliability in distribution systems are described. A mathematical representation of the distribution system is constructed by evaluating the components and the relation between them. %The approaches build a mathematical representation of the distribution system by evaluating the systems components and the relation between them. 
To capture the dynamics in the system behavior, Markov models can be applied. From the state models, both transient and stationary properties can be extracted, such as the availability and the reliability of the system. 
%One approach is to use Markov techniques in combination with state-space diagrams. Through the state-space diagrams, the state probability can be conducted which can be used to evaluate the reliability of the system. 
To capture the structure and the interrelation between the system components, Markov models are not scalable. Alternatively, the system can be represented by a structural model (such as Reliability Block Diagrams and Fault trees), from which 
%Another solution is to approximate the system by building 
minimal cut sets %of the system components illustrating their relation and dependency. 
can be defined and stationary system properties can determined.
Another analytical approach, RELRAD, uses the fault contribution from all the network components to calculate the individual load point reliability in the system \cite{RELRAD}. However, these methods consider the passive operation of the network and are therefore not optimal when analyzing modern distribution systems. 

%This can be conducted by making state-space diagrams of the possible states of the system components. The state-space diagrams will then give the probability of being in the different states. Markov techniques can then be used to evaluate the reliability of the system. 
 %can be used to evaluate the reliabiliyt of the system. This result in a 
 %
Through simulation, a more accurate representation of a modern distribution system is possible. In a simulation, information about the variability of the reliability indices is included.  
% more accurate representation of the system will be given and information about the variability of the reliability indices and significant outcomes can be given.
Monte Carlo simulation is the approach  most  frequently  used  for  studying  combined ICT-power system \cite{chaudonneret_representation_2012}. Multiple studies have used Monte Carlo simulation as an approach to evaluate reliability in modern distribution systems with ICT and active components such as distributed generators (DG). In \cite{celli2013reliability}, a comprehensive study is conducted on modeling smart distribution networks with ICT and DGs. The reliability of a distribution system with the possibility of creating intentional islands with DGs is evaluated in \cite{de2017reliability} where a sequential Monte Carlo method is used to evaluate the reliability. Whereas in \cite{borges2012overview}, an overview of reliability models and methods for distribution systems with renewable DGs is presented with the focus on the impact from the DGs. Here, Monte Carlo is often the preferred method. However, even though the studies illustrate comprehensive methods, they do not bring forward a general tool for the reliability assessment of modern distribution systems. In \cite{escalera2018survey}, a survey of reliability assessment techniques for modern distribution systems is conducted. The research points out Monte Carlo simulation as a commonly used strategy. This is mostly due to its stochastic nature which allows for more accurate modeling of the system where the complexity and uncertainty of the system are considered. The paper also points out the need for new reliability assessment tools for active distribution systems.

This paper aims to introduce a tool for reliability assessment in a smart and active power system. The tool, RELSAD (RELiability tool for Smart and Active Distribution networks), offers the following features: 
\begin{itemize}
    \item A foundation for calculating the reliability of modern distribution systems with smart and active components. 
    \item A simulation tool that includes active participation of different power sources such as distributed generation, microgrids, and batteries.
    \item A simulation tool that includes ICT components and the dependency between the ICT components and the power system.  
\end{itemize}
The tool is made as a Python package openly available at Github with package documentation \cite{RELSAD}.
The use of the tool is demonstrated by examples performed on the IEEE 33-bus network \cite{baran1989network} including active and smart components.  

The rest of this paper is organized as follows; In Section \ref{structure}, the package structure is described, and the system components are presented. Section \ref{function} aims to describe the core functionality of the method. A case study is illustrated in Section \ref{casestudy} before the results are presented and discussed in Section \ref{discussion}. The conclusions are presented in Section \ref{conclusion} with future work. 

%% file: Sections/02SystemStructure.tex
\section{Package Structure}
\label{structure}

The structure of the program is divided into an electrical power system part and an ICT system part. This chapter aims to describe both system structures. 

\subsection{Electrical power system}

\begin{figure}[tb]
    \centering
    \begin{subfigure}[b]{0.5\textwidth}
        \centering
        \includegraphics[width=\linewidth]{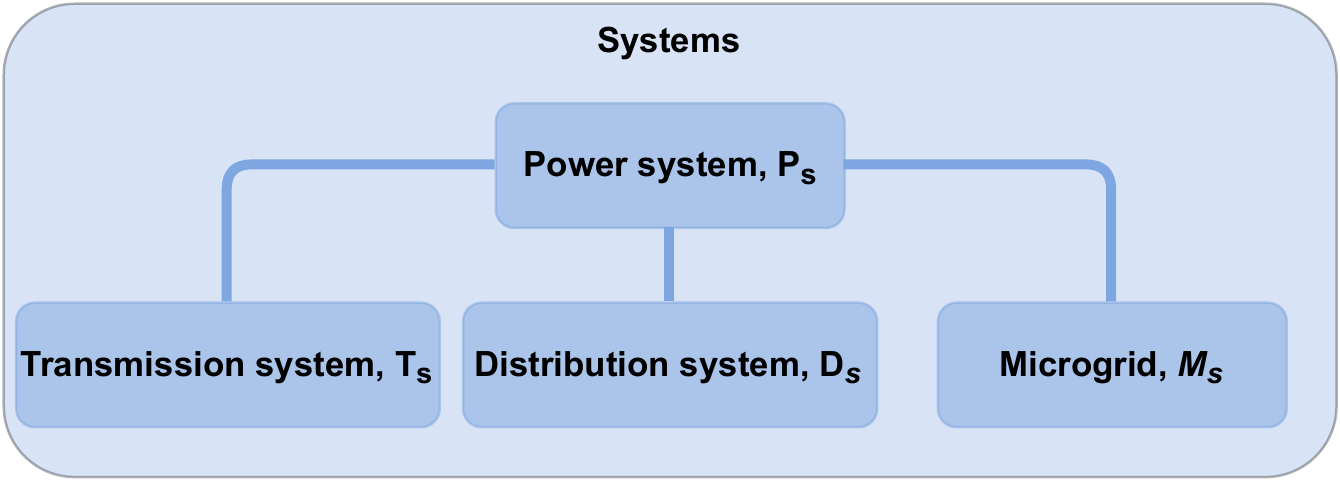}
        \caption{The system structure of the power system networks}
        \label{fig:ElStructureSystems} 
     \end{subfigure}
     
     \begin{subfigure}[b]{0.5\textwidth}
        \centering
        \includegraphics[width=\linewidth]{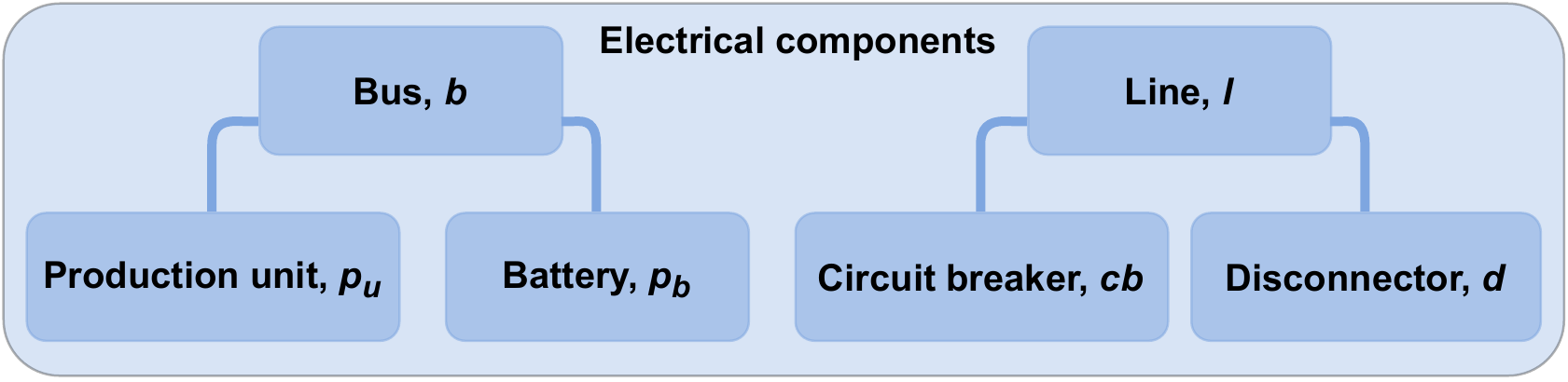}
        \caption{The components structure of the power system components}
        \label{fig:ElStructureComponents}
     \end{subfigure}
  %\label{fig:ComStruct}
  \caption{The electrical power system structure}
\end{figure}

The modeling of the electrical power system is divided into two categories, 1) \textit{Systems} and 2) \textit{Electrical components}. Fig. \ref{fig:ElStructureSystems} and \ref{fig:ElStructureComponents} illustrates the structure of the systems and electrical components respectively. 

The systems are based on a parent power system - $P_s$, which further is built upon a combination of one or more transmission systems - $T_s$,  distribution systems - $D_s$, and microgrids - $M_s$. The transmission network is not yet implemented and is therefore not discussed in this paper. 

Furthermore, the power systems are built up of different electrical components such as lines - $l$, buses - $b$, disconnectors - $d$, circuit breakers - $cb$, loads - $p_d$, production units - $p_u$, and batteries - $p_b$. Together, both the implemented systems and the electrical components in the systems constitute the power system. The load is an attribute of a bus. 

%The electrical power system, $P_s$, is structured into a combination of transmission systems, $T_s$, and distribution systems, $D_s$. However, the transmission network is not implemented and will therefor not be included in this paper. The power system is built up of the electrical power system components lines, $l$, buses, $b$, disconnectors, $d$, circuit breaker, $cb$, load, $p_d$, and production units, $p_u$. The failures are implemented 

\subsection{ICT system}

The ICT system, $I_s$, is built up of three different communication components; 
\begin{enumerate}
    \item \textbf{Controller}, $dc$: The software containing the algorithm such as SCADA system. The data center is responsible for monitoring and computations in the system. Functions as the controller of the system. 
    \item \textbf{Sensor}, $s$: Such as \textit{intelligent electronic device} (IED) placed on the system lines. Sensors can send line data information to the controller. The data can be used to, e.g., find out which line is faulty.
    \item \textbf{Intelligent switch}, $d_i$: An Intelligent switch that can receive commands about the switch position from the controller and automatically open/close the breaker. 
\end{enumerate}
%a data center, $dc$, which is the software such as SCADA that manage and does all the computations of the system, intelligent sensor, $s$, on the lines in the system that can send information about the line data to the software, and smart disconnectors, $d_i$, which can receive information and commands about the switch position from the software. 

\begin{figure}[tb]
    \centering
    \includegraphics[width=0.4\textwidth]{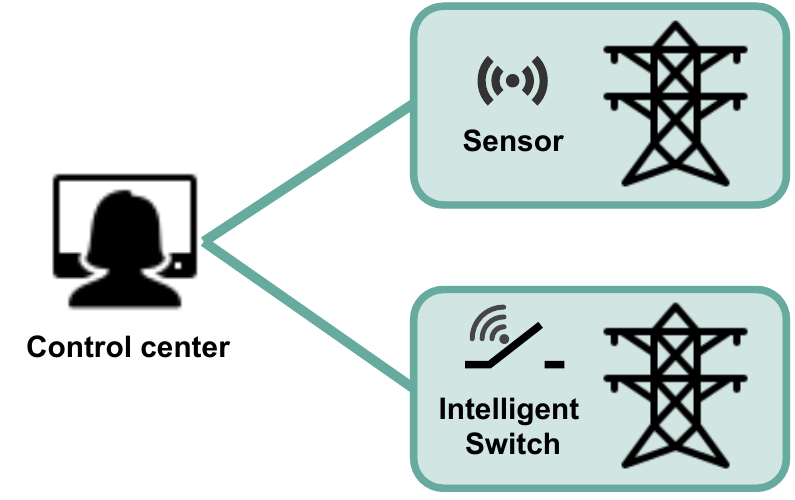}
  \caption{The communication component structure}
  \label{fig:ComStruct}
\end{figure}

The ICT components are placed inside the given power system the component belongs to ($D_s$ or $M_s$), e.g., an intelligent switch is connected to a disconnector in a distribution system. 
Fig. \ref{fig:ComStruct} illustrates the setup and communication between the different ICT components. The sensor can send information about the line to the controller whereas the controller can command the intelligent switch to open or close a switch automatically remotely. %The smart switch can open and close the disconnectors and the circuit breakers remote and dose then not need physically. 

\subsection{Class structure}

\begin{figure*}[t]
    \centering
    \includegraphics[width=\textwidth]{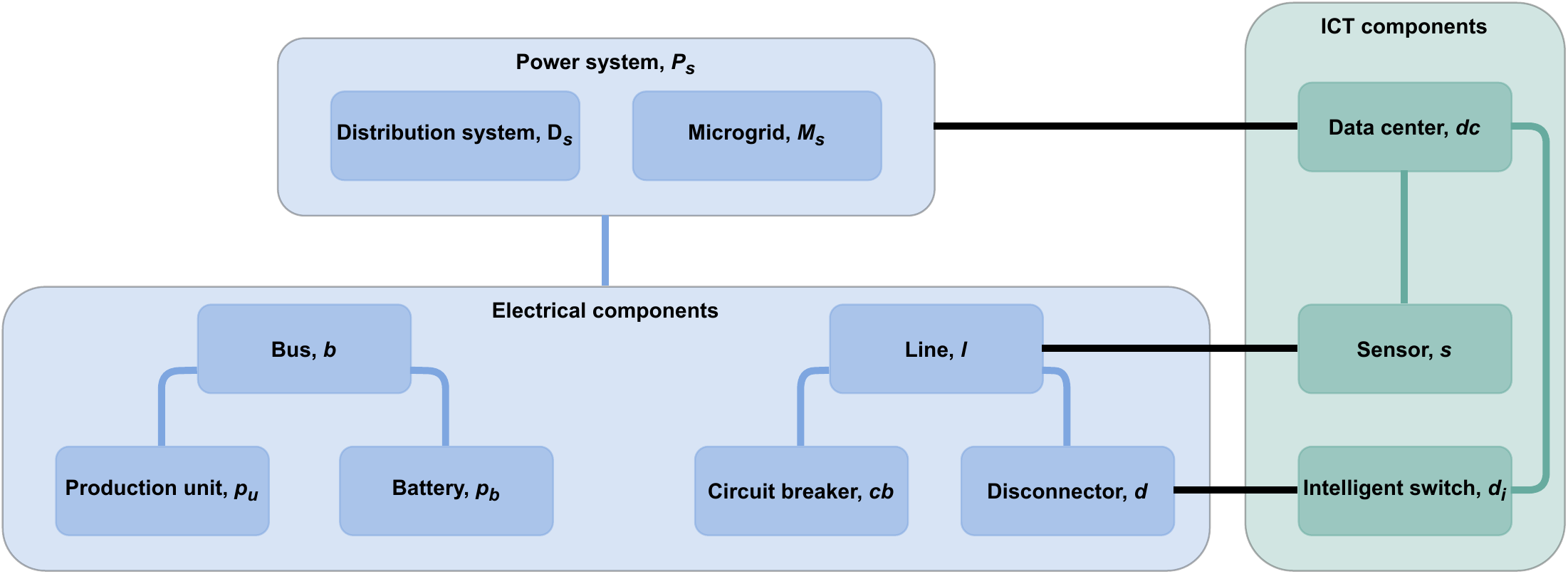}
  \caption{The component structure}
  \label{fig:CompStructure}
\end{figure*}

The reliability assessment tool is constructed based on an object-oriented programming approach with a class structure for the different network components. The class structure is divided between an electrical power network structure and an ICT network structure, where the logic inside and between the different network structures are implemented in the program. Fig. \ref{fig:CompStructure} illustrates the class structure based on the layers in the system as well as the connection between the different classes. The electrical power system classes are illustrated in blue, while the ICT component classes are illustrated in green. First, a $P_s$ is created before a combination of one or multiple $D_s$ and $M_s$. After that, all the different components, both electrical and ICT, can be created and added to the associated parts of the system. 
%components in the system where it belongs. 

%Fig. \ref{fig:ClassStructure} gives an overview of the class structure in the model. In the figure, some important class attributes are included to illustrate how the component class is built up. 

%% file: Sections/03SystemFunction.tex
\section{Core functionality}
\label{function}

\subsection{Monte Carlo simulation and incremental procedure}

RELSAD calculates the reliability of a power system through Monte Carlo simulations.
%The reliability of a power system in RELSAD is calculated through Monte Carlo simulations. 
The Monte Carlo method is a stochastic simulation approach based on a random sampling of input variables to solve numerical problems such as stochastic problems. Monte Carlo uses randomness to compute the statistical distributions of parameters of interest. In the simulation, time steps are used to comprehend the behavior of the power system to give a better representation of the actual system. In RELSAD, a sequential simulation model is developed with a user-chosen increment interval.  %By using simulation, the behavior of the system is easier to measure since the simulation allows for using time-steps that can comprehend different system behaviors which will lead to a better representation of the actual system. 

First, $P_s$ is created and initialized with all the systems and components in the $P_s$. Algorithm \ref{alg:procedure} illustrates the incremental procedure of the model. The algorithm will first set the load and the production at the nodes in the system before it will draw the status of the components in the $P_s$. The algorithm will then check for failures in the $P_s$. If there are any failures, the $D_s$ and/or $M_s$ with failed components will be divided into sub-systems and a load flow and minimization problem will be solved for each sub-system (see section \ref{sec:loadflow}). After this, the systems and components' history variables are updated. The history variables are further used to evaluate the reliability for the whole system, $P_s$, the individual $D_s$ and $M_s$, and the individual load points. The rest of this section will explain the different procedures of the reliability evaluation in greater detail. 
%To consider the function of the system, the state of the system at the simulated time-step needs to be calculated. This can be done through load flow calculation. By adding load flow calculation into the simulation, the condition and behavior of the system can be calculated at all the wanted time-steps. In addition, the behavior of the battery and the DGs will, therefore, also be included and considered and the impact from the resources in the microgrid can be calculated more accurately. The electrical consequence of a failure can be calculated and varied load and production can be simulated. 
\begin{algorithm}
    \SetAlgoLined
     Set bus $p_d$ and $p_u$\;
     Draw component fail status\;
     \If{Failure in $P_s$ (in systems, $D_s$ and $M_s$)}{
         Find sub-systems\;
         \ForEach{sub-system}{
            Update $p_b$ demand (charge or discharge rate)\;
            Run load flow\;
            Run optimization problem\;
         }
         Update history variables\;
     }
    \caption{Increment procedure}
    \label{alg:procedure}
\end{algorithm}
    
\subsection{Failure mode}
\begin{figure}[tb]
    \centering
    \includegraphics[width=0.35\textwidth]{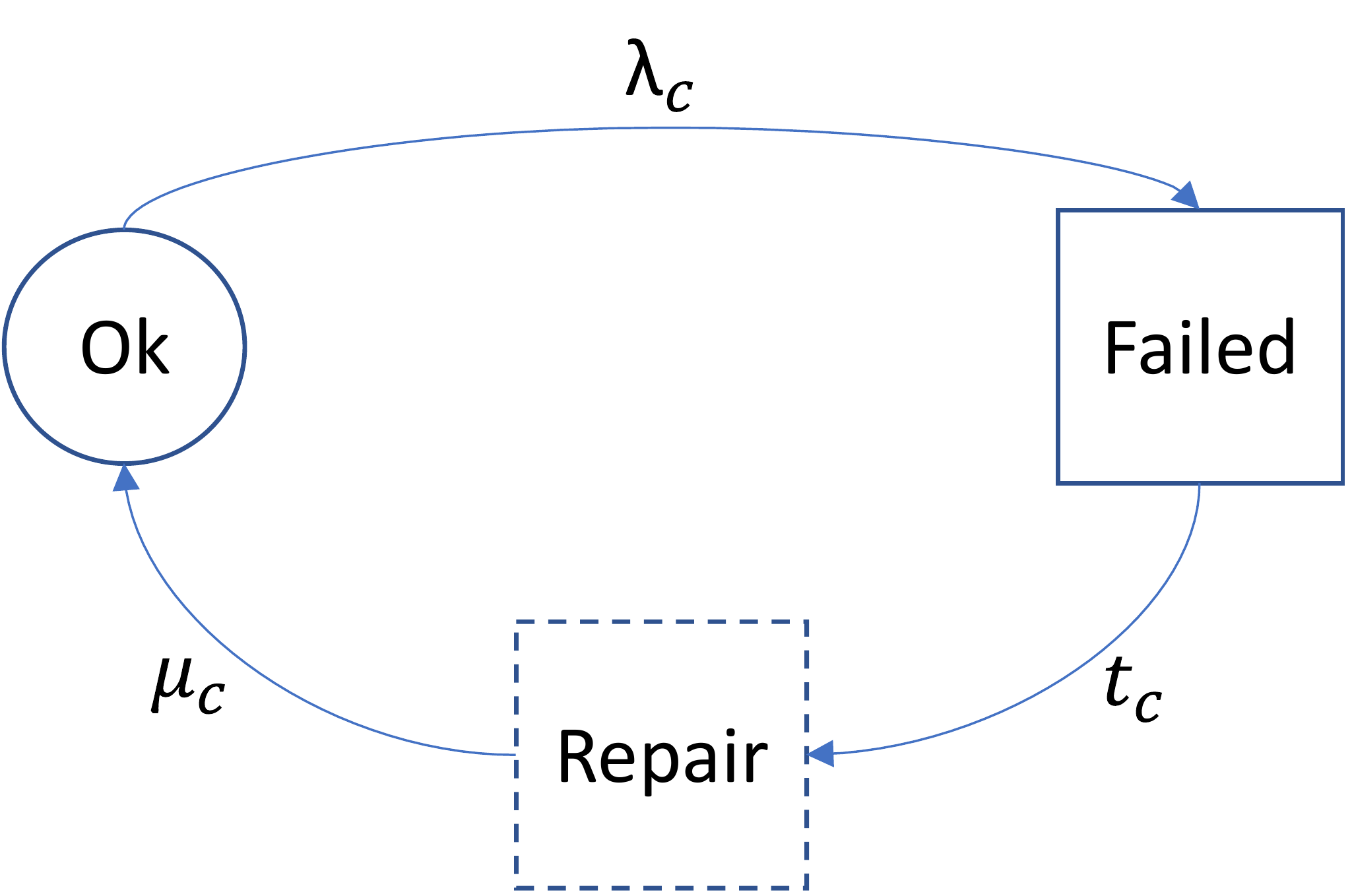}
  \caption{A simple Markov model of the system components. $\lambda_c$ is the failure rate of component $c$, $\mu_c$ is the repair rate of component $c$, and $t_c$ is the time from component failure until the component is under repair.}
  \label{fig:Failmode}
\end{figure}

The system is constructed with a user-specified failure rate and repair time for the different system components. Fig \ref{fig:Failmode} illustrates a simple Markov model of the system components. The system components are, at the moment, implemented with three different possible states, a working state when the component is up and running, a failed state where the component is down and not working, and a repair state where the component is under repair. The failed state for the controller is further divided into two different failure modes, hardware failures, and software failures. The repair state is a sub-state of the failed state, but the difference is that the failed component is acknowledged and is under repair. 

The transition between the working state and the failed state is decided based on the failure rate of the component and is drawn by the simulator for each increment. If the component is in a failed state, a new state for the component will not be drawn until the component is repaired. 
\begin{algorithm}
    \SetAlgoLined
     \If{Failure on $s$ or \textit{software failure} on $dc$}{
        repair\_time = new\_signal\_time\;
        Draw signal success\;
        \uIf{Component new signal received}{
        \Return{repair\_time} \;
        }
        \Else{
            repair\_time += reboot\_time\;
            Draw reboot success\;
            \uIf{Reboot successful}{
                \Return{repair\_time}\;
            }
            \Else{
                Start manual repair\;
                \Return{repair\_time}\;
            }
        }
     }
    \caption{Repair procedure}
    \label{alg:Repairmode}
\end{algorithm}

The transition between the failed state and the repair state can vary based on the time it takes for the system to discover the failure on the failed component. E.g., for an intelligent switch, the failure will not be discovered until the switch is requested to open or close. The same applies to the sensor, where the failure is not discovered until the sensor is called upon. For a failure on a line, the $t_c$ is decided by the \textit{sectioning time} of the system, i.e., the time it takes for the system to find and isolate the failed section in the power system, and the \textit{sectioning time} is initialized by the opening of the circuit breaker. 

The transition between the repair state and the working state is determined by the outage or repair time of the component. For the sensor and the controller during software failures, the repair time follows a multi-phase recovery as seen in Algorithm \ref{alg:Repairmode}. This will result in three different possible outcomes related to the time it takes for the system to address the failure in the power system. First, a new signal is searched for. If that fails, the component is rebooted. If the reboot fails to fix the problem, the component needs manual repair. 

\subsection{Load flow}
\label{sec:loadflow}

To consider the behavior of the system, the state of the system at the simulated time-step needs to be calculated. This can be done by adding load flow calculations into the simulation, calculating the condition and behavior of the system at all the specified time steps. 
%This can be done through load flow calculation. By adding load flow calculation into the simulation, the condition and behavior of the system can be calculated at all the wanted time-steps. 
In addition, the behavior of the battery, the DGs and the other flexible resources are included and an accurate calculation of the impact these sources have on the $P_s$ can be evaluated. The electrical consequence of a failure can be calculated and changed load and production can be simulated.  

%and considered and the impact from the resources in the microgrid can be calculated more accurately. The electrical consequence of a failure can be calculated and varied load and production can be simulated. 

A load flow solver is implemented in the tool to capture the behavior of the system. Since the system includes smart and active components, which can restore some parts of the network, load flow calculations are added to determine the electrical consequence of an outage in the system. 
Since the model currently only considers radially operated networks, the chosen load flow solver is based on a Forward-Backward Sweep (FBS) approach. The FBS will calculate the load flow parameters of the network iteratively in a backward and forward sweep order until convergence is achieved \cite{LoadFlow, haque1996load}. The FBS is preferred as the load flow solver for radially operated networks since the FBS approach does not go through the Jacobian matrix, as the Newton-Raphson method, which for some weak networks can cause an ill-conditioned matrix and result in convergence problems for the load flow calculation. The FBS approach is based on PyDSAL \cite{fosso2020pydsal}. 

In addition to the load flow solver, a optimization problem is implemented with the objective to determine the amount of shedded load in the network that will minimize the \textit{Cost of Energy not Supplied} (CENS).
This is subjected to load flow balance and limitations in the network. The minimization problem can be seen in eq. \ref{eq:minproblem}. Here $C_n$ is the cost of shedding load at node n while $P_{n}^{s}$ is the amount of shedded power at node n. $P_{j}^{g}$ is the production from generator $j$. $P_{k}^{d}$ is the load demand at node $k$ while $P_{i}^{l}$ is the power transferred over line $i$. $\gamma_i= 1$ if line $i$ is the starting point, -1 if line $i$ is the ending point of the line. $\lambda_j= 1$ if there is a generation unit at node $j$, 0 else. $\mu_k = 1$ if there is a load on node $k$, and $\mu_k = 0$, otherwise.
\begin{align}
    &\underset{P^{s}_{n}}{\text{minimize}}
    \quad \mathcal{P}_s = \sum_{n = 1}^{N_n}   C_{n}\cdot P^{s}_{n} \label{eq:minproblem} \\
    &\text{subject to: } \nonumber \\
    &\begin{aligned}
        \sum_{i=1}^{N_l} \gamma_i \cdot P^{l}_{i} &= \sum_{j=1}^{N_g} \lambda_j \cdot P^{g}_{j} - \sum_{k = 1}^{N_n}& \mu_k& \cdot (P^{d}_{k} - P^{s}_{k})\\
        \min P_{j}^{g} &\leq P_{j}^{g} \leq \max P_{j}^{g} &\forall j&=1,\dots,N_{g}\\
        0 &\leq P_{k}^{s} \leq P_{k}^{d}  &\forall k&=1,\dots,N_{n}\\
        \left| P_{i}^{l} \right| &\leq \max P_{i}^{l}  &\forall i&=1,\dots,N_{l}\\
    \end{aligned} \nonumber
\end{align}

\subsection{Reliability indices}

To evaluate the reliability of the $P_s$ and the systems contained in $P_s$, some reliability indices are calculated during the increment procedure. These are updated in the history variables of the different components and systems in $P_s$. 

The considered reliability indices calculated in RELSAD are divided into two different indices groups, 1) \textit{Load- and generation-oriented indices} and 2) \textit{Customer-oriented indices}. 

\subsubsection*{\bf Load- and generation-oriented indices}
The load- and generation-oriented indices aim to indicate the electrical consequence of failures in the system. 
The energy not supplied in eq. \ref{eq:ENS} indicates how much power is not being able to be served where $U_s$ is the outage time of load point $i$ and $P_i$ is the load at load point $i$.

\begin{equation}
    \label{eq:ENS}
    {\tt ENS}_{s} = \sum {U_{i}P_{i}}
\end{equation}

%\begin{equation}
 %   \label{eq:Pinterrupted}
%    P_{interr,s} = \lambda_{s}P_{s}
  %   \mathcal{P}_{I} = \lambda_{s}P_{s}
%\end{equation}

The interruption cost for the system can be calculated as seen in eq. \ref{eq:cost} \cite{943073}, where $c_i$ is the specific interruption cost for each customer category at load point $i$. %In this paper, the cost for interruption is based on the KILE-cost. 

\begin{equation}
    \label{eq:cost}
    {\tt CENS_{s}} = \sum {\tt ENS}_{i}c_{i}
\end{equation}

\subsubsection*{\bf Customer-oriented indices}

The customer-oriented indices aim to indicate the reliability of the distribution system by the interruption experienced by the customers. There are multiple different types of indices, but in this paper, only three important indices are investigated, 

\begin{enumerate}
    \item System Average Interruption Frequency Index% (SAIFI)
\begin{equation}
    \label{eq:SAIFI}
    {\tt SAIFI} = \frac{\sum_{\forall i} \lambda_{i}N_{i}}{\sum N_{i}}
\end{equation}
where $N_{i}$ is the {\em total number of customers served}, and $\sum_{\forall i} \lambda_{i}N_{i}$ is the {\em total number of customer interruptions}. ${\tt SAIFI}$ is a measure of the frequency of interruptions the customers in the system expect to experience. 
\item System Average Interruption Duration Index% (SAIDI),
\begin{equation}
    \label{eq:SAIDI}
    {\tt SAIDI} = \frac{\sum U_{i}N_{i}}{\sum N_{i}}
\end{equation}
where $\sum_{\forall i} U_{i}N_{i}$ is {\em total number of customer interruption} durations. ${\tt SAIDI}$ is a measure of the expected duration of interruptions a customer is expected to experience. 
\item Customer Average Interruption Duration Index% (CAIDI). 
\begin{equation}
    \label{eq:CAIDI}
    {\tt CAIDI} = \frac{\sum U_{i}N_{i}}{\sum_{\forall i} \lambda_i N_{i}} = \frac{\tt SAIDI}{\tt SAIFI}
\end{equation}
${\tt CAIDI}$ is the ratio between ${\tt SAIDI}$ and ${\tt SAIFI}$ and measures the average duration each given customer in the system is expected to experience. 
\end{enumerate}

%% file: Sections/04UserInterface.tex
\section{User interface}

\subsection{Networks}
The user will have the opportunity to create the network(s) to be investigated. The network should be made in Python as a \textit{.py} file. The user creates a $P_s$ before the different system components. The network is created by connecting all the elements. %An example of creating a simple two-bus network is seen below. 
For more detailed information, see the documentation in the package repository at Github \cite{RELSAD}.

%Du lager nettverk ved å opprette komponentobjekter, koble disse sammen, og legge de til i et nettverksobjekt. 

% \inputminted[
% frame=lines,
% framesep=2mm,
% baselinestretch=1.2,
% fontsize=\footnotesize,
% linenos]{python}{Figures/Test_PSCC.py}
% %\lstinputlisting[language=Python]{Figures/Test_PSCC.py}

\subsection{Data input}

After the $P_s$ is created, a \textit{run file} with the Monte Carlo commands needs to be established. Here, the load and generation at the buses are set.
Data input such as system load and production should be collected through a self-made Python file (.py file), that can read and fetch the system data.  
In the \textit{run file}, the data is included and connected to the right bus and generation element. The load and production data are included through Python dictionaries made in the \textit{run file}. 
To include cost function related to the different load types, these should be included in the \textit{run file} as a part of the \textit{load dictionary}. 

Since the simulator allows for user-chosen time steps, a function is included in the tool to interpolate the data for the specified time step.

%\inputminted[
%frame=lines,
%framesep=2mm,
%baselinestretch=1.2,
%fontsize=\footnotesize,
%linenos]{python}{Figures/Monte_Carlo_Test.py}

%Last og produksjonsdata inkluderes i simuleringen gjennom python dictonaries. Brukeren selv oppretter foreksempel en python som inneholder eller kan lese inn data også legges dataen til tilhørende komponeter i systemet og last- og produksjons dictionaries opprettes for nettet. 

%Interpoleringen som intepolerer etter inkrementstørrelse. 

\subsection{Simulation increment}

RELSAD has a user-defined time interval, meaning that the user can specify the simulation increment. The system is built with a \textit{TIME} utils class which will convert all the time units used in the system to a common time unit. 
Be aware that the reliability evaluation will be less accurate with coarser time steps, but with lower simulation time compared to a higher resolved time step. 
In addition, the tool is implemented with the possibility of parallel processing, which will speed up the simulation time. 

%% file: Sections/06CaseStudy.tex
\section{Case Study}
\label{casestudy}

\subsection{Test network}

\begin{figure*}[tb]
    \centering
    \includegraphics[width=\textwidth]{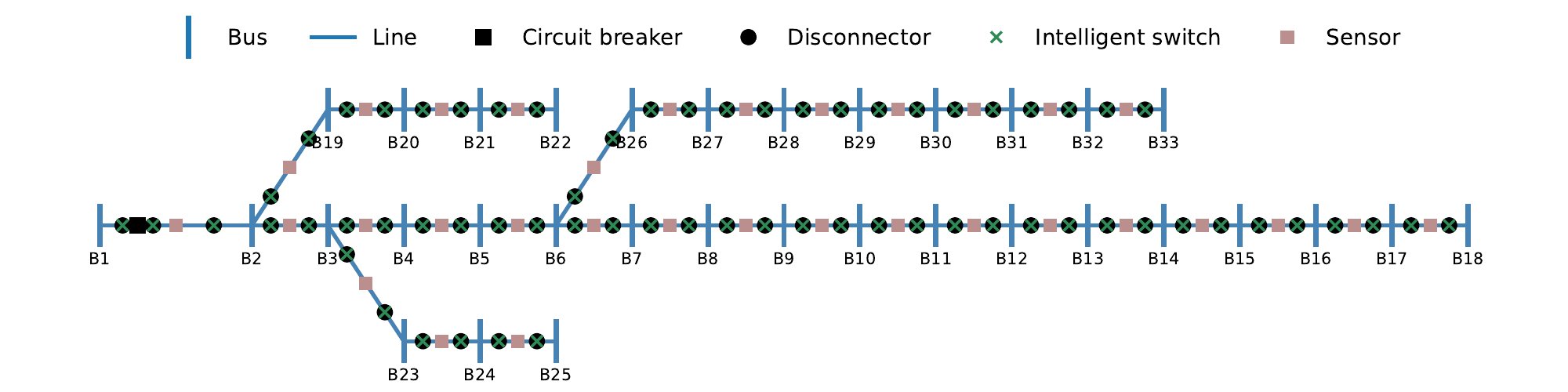}
  \caption{System topology with ICT}
  \label{fig:SystemTopology}
\end{figure*}

The model is applied on the IEEE 33-bus system seen in Fig. \ref{fig:SystemTopology}.
The distribution system includes the discussed ICT components, a battery placed on B30, and a wind power plant placed on B15.

The network data and the reliability data can be seen in Tab. \ref{tab:MicrogridSpec} and Tab. \ref{tab:SystemSpec} respectively.
The increment is set to one hour. The load data is generated based on the FASIT requirement specification which has hourly load increments and different load types \cite{FASIT}. The weather data for the load profiles and the wind turbine are collected from a location east in Norway. 
%The system includes two batteries, one in the microgrid and one placed in the distribution system. The specifications for the load, production units, and the batteries can be seen in Tab. \ref{tab:SystemSpec}. 

\begin{table}[H]
    \caption{Distribution system specifications}
    \center
\begin{tabular}{cc} 
     \hline
     \textbf{Component} & \textbf{Specification} \\ \hline
     Battery &  \makecell[l]{Max capacity: 1 MWh \\ Inverter capacity: 500 kW \\ Min SOC: 0.1 } \\  \hline
    %Solar power %& Number of modules: 5000 \\ 
    %  \cline{2-2}
     %& Max power: $\sim$ 1 MW \\ \hline
     Wind power %& Number of turbines: 2 \\ 
     %\cline{2-2}
     & Max power: $\sim$ 5 MW \\
     Load distribution system & Peak load $\sim$ 6.3 MWh  \\ \hline
\end{tabular}
\label{tab:MicrogridSpec}
\end{table}

\begin{table}[H]
    \caption{Failure rate and repair time of the system components. The failure rate and repair time for the power system components are based on interruption statistic from the Norwegian power system, while the failure and repair rates for the ICT components are based on \cite{amare2018modeling}.}
    \center
    \resizebox{\columnwidth}{!}{
    \begin{tabular}{p{0.40\columnwidth}p{0.26\columnwidth}p{0.26\columnwidth}} 
     \hline
     \textbf{Component} & \textbf{Failure rate [failure/year]} & \textbf{Outage time/repair time [h/failure]} \\ \hline
     Line &  0.07 & 4 \\ 
     Transformer & 0.007 & 8 \\ 
     Intelligent switch & 0.03 & 2\\ 
     Sensor & 0.023 & \makecell[l]{New signal: 2 sec \\ Reboot: 5 min \\ Manual repair: 2} \\ 
     Controller (hardware failure) & 0.2 & 2.5 \\ 
     Controller (software failure) & 12 & \makecell[l]{New signal: 2 sec \\ Reboot: 5 min \\ Manual repair: 0.3} \\ \hline
\end{tabular}}
\label{tab:SystemSpec}
\end{table}

\subsubsection{Battery strategy}
The battery is assumed to follow the electrical market in the system, meaning that the batteries will sell and store energy based on market prices. This will be simulated by predicting the SOC level of the battery following a uniform distribution between max and min SOC.

\subsection{Scenario description}

The case study aims to illustrate how the presented model is working and how it calculates the reliability in a more advanced power system. 
The case study consists of four different scenarios seen below: 

\begin{enumerate}
    \item \textbf{No ICT, storage, and generation in the distribution network:} The network is operated traditionally without any ICT and production units present. 
    \item \textbf{Distribution system with storage and generation, no ICT:} Including the production units, a battery, and wind turbines, in the system. The system can now restore supply to some parts of the system.
    \item \textbf{Distribution system with ICT, no storage and generation:} Including the ICT components, sensor, controller, and intelligent switch. No production units are present. Fast sectioning time in the system when the ICT components are working. 
    \item \textbf{Distribution system with ICT, storage and generation:} The distribution system includes ICT components, storage, and generation.
\end{enumerate}

By using hourly increments in the simulation, some information will be inaccurate. The time step is set to one hour to follow the manual sectioning time of the system. However, if there are ICT components in the system, the sectioning time will in most of the cases be very small (seconds or minutes during some failure events), the shedded electrical power in these time intervals will therefore not be accounted for in an hourly increment simulation and the result is, therefore, a little optimistic for the cases with ICT.

%% file: Sections/07Discussion.tex
\section{Result and discussion}
\label{discussion}

\subsection{Result}
The network is simulated for the different cases with a time interval of 1 hour. The result for the simulations is illustrated in Tab. \ref{tab:Result}. As expected, the $\tt ENS$ will decrease for the cases with ICT and generation units compared to the base case. The largest difference can be seen for the cases with ICT since the sectioning time is very low compared to the no-ICT cases. This will lead to fewer load points experiencing shedding and will also result in a lower $\tt SAIFI$. The case with only generation and battery will give lower $\tt ENS$ compared to base cases since some parts of the network will gain supply. 

In addition, $\tt SAIDI$ is decreased for all the cases compared to the base case since the load points will experience a decreased period of shedded load. However, $\tt CAIDI$, the expected outage duration for the customer, will increase. In the cases without ICT, all the load points will experience an outage during the sectioning time of the system forcing $\tt CAIDI$ closer to one which is the period of the sectioning time. For the cases with ICT however, only the customers that are directly affected by the failure will experience an outage in most of the cases. This will result in a much lower $\tt SAIFI$, but $\tt SAIDI$ will not decrease that much since most of the outage time in the system is related to the outage time of the system components and not the sectioning time. 

Another factor to consider is the simulation interval. Since the simulation interval is set to 1 hour, the small periods where the ICT system is searching for a new signal and rebooting are neglected. This means that the result for the ICT cases is a bit optimistic compared to a real situation. In addition, the result for $\tt CAIDI$ will be off since the small sectioning time that will occur in the cases with ICT is not considered. It is therefore important to consider simulation time up against the measure of the accuracy of the result. A higher time resolution on the simulation will give longer simulation time, but will also give a more accurate result. In this paper, the aim is to introduce RELSAD, and a simulation time of 1 hour is acceptable for this analysis. However, this is an important factor to address and consider when building simulation tools. A possible solution is to have a function that can create a higher time resolution on the simulation interval when this is needed during outages in the system. The simulation interval during the outage will then be set based on the situation of the components in the system, but this is not yet implemented.

\begin{table}[H]
    \caption{Reliability indices of the four cases}
    \center
    \resizebox{\columnwidth}{!}{
    \begin{tabular}{p{0.25\columnwidth}p{0.25\columnwidth}p{0.25\columnwidth}p{0.25\columnwidth}}
        \hline 
        &&&\\ 
     \multicolumn{4}{c}{\textbf{Case 1: No ICT, storage and generation in the distribution network}} \\
     &&&\\
     \textbf{ENS} & \textbf{SAIDI} & \textbf{SAIFI} & \textbf{CAIDI} \\
     39.8309 &  9.9317 & 5.4205 & 1.8322 \\ \hline
     &&&\\
     \multicolumn{4}{c}{\textbf{Case 2: Distribution system with storage and generation, no ICT}} \\
     &&&\\
     \textbf{ENS} & \textbf{SAIDI} & \textbf{SAIFI} & \textbf{CAIDI} \\
     34.8709 & 9.8626 & 5.3633 & 1.8389 \\ \hline
     &&&\\
     \multicolumn{4}{c}{\textbf{Case 3: Distribution system with ICT, no storage and generation}} \\
     &&&\\
     \textbf{ENS} & \textbf{SAIDI} & \textbf{SAIFI} & \textbf{CAIDI} \\
     16.6798 &  7.6166 & 1.0915 & 6.9781 \\ \hline
     &&&\\
     \multicolumn{4}{c}{\textbf{Case 4: Distribution system with ICT, storage, and generation}} \\
     &&&\\
     \textbf{ENS} & \textbf{SAIDI} & \textbf{SAIFI} & \textbf{CAIDI} \\
     14.6094 &  7.5903 & 1.0894 & 6.9674 \\ \hline
\end{tabular}}
\label{tab:Result}
\end{table}

%% file: Sections/08Conclusion.tex
\section{Conclusion and future work} 
\label{conclusion}

This paper has introduced RELSAD as a reliability assessment tool for the future smart and active distribution system. RELSAD is built up as a Python package based on object-oriented programming and will be published as open-source for public use upon publication of the paper. RELSAD introduce a simulation method that includes DGs, batteries, microgrids, and ICT components into the reliability evaluation of modern distribution network. By this, it can comprehend the new components and technology and evaluate the reliability of a modern distribution system.

One important factor to consider when using a simulation tool is the time-step of the simulation. A higher time resolution on the time-step gives a more accurate result, but the simulation time will increase. Therefore, it is important to determine the required accuracy of a given problem. 

%determine the measure of the accuracy of a given problem.  

\subsection{Future work}

The tool will be further developed, tested, documented, and made more robust. In addition, some features will be addressed in more detail 

\begin{enumerate}
    \item Further development of the ICT components by including more failure modes.
    \item Investigate possible solutions for more precise analysis of the result by having more resolved time intervals or a more precise increment timeline. 
    \item Include additional features such as the inclusion of more ICT components and the ICT network. 
\end{enumerate}

%% file: Sections/09Acknowledgement.tex
\section{Acknowledgement}
\label{acknowledgement}

This work is funded by CINELDI - Centre for intelligent electricity distribution, an 8-year Research Centre under the FME-scheme (Centre for Environment-friendly Energy Research, 257626/E20). The authors gratefully acknowledge the financial support from the Research Council of Norway and the CINELDI partners. The authors would like to thank Jonas Rudshaug for his great technical support.